\documentclass[english,5p]{elsarticle}
\usepackage[T1]{fontenc}
\usepackage[latin9]{inputenc}
\usepackage{amsmath}
\usepackage{cancel}
\usepackage{graphicx}
\usepackage{babel}

\begin{document}

\begin{frontmatter}{}

\title{Localization in Coupled Finite Vibro-Impact Chains: Discrete Breathers
and Multibreathers}

\author[rvt]{I. Grinberg\corref{cor1}}

\ead{GrinbergItay@gmail.com}

\author[rvt]{O.V. Gendelman}

\ead{ovgend@technion.ac.il}

\cortext[cor1]{Corresponding author}

\address[rvt]{Faculty of Mechanical Engineering, Technion \textendash{} Israel
Institute of Technology, Haifa, Israel}
\begin{abstract}
We examine the dynamics of strongly localized periodic solutions (discrete
breathers) in two-dimensional array of coupled finite one-dimensional chains of oscillators.
Localization patterns with both single and multiple localization sites
(multibreathers) are considered. The model is scalar, i.e. each particle can move only parallel to the axis of the chain it belongs to. 
The model involves symmetric parabolic
on-site potential with rigid constraints (the displacement domain
of each particle is finite) and a linear nearest-neighbor coupling
in the chain, and also between the neighbors in adjacent chains. When
the particle approaches the constraint, it undergoes an elastic Newtonian
impact. The rigid impact constraints are the only source of nonlinearity
in the system. The model allows easy computation of highly accurate approximate solutions for
the breathers and multibreathers with an arbitrary set of localization
sites in conservative setting. The vibro-impact nonlinearity permits
explicit derivation of a monodromy matrix for the breather and multi-breather solutions.
Consequently, the stability of the derived breather and multibreather
solutions can be studied in the framework of simple methods of linear
algebra, without additional approximations. It is shown that due to
the coupling of the chains, the breather solutions can undergo the symmetry breaking. 
\end{abstract}

\end{frontmatter}{}

Localization in discrete dynamical systems with nonlinearity has been
of growing interest in recent years\cite{Flach1998,Flach2008,Vakakis1996,Campbell2004,Vakakis2008,Anderson1958,PIERRE1987549,Bendiksen2000,Cai1994}.
Nonlinearity enables localization even in a purely homogeneous system,
without the need for a disorder as is the case of ordinary linear systems.
The primary phenomenon discussed in this work is the Discrete Breather
(DB), often referred to as intrinsic localized mode (ILM) or discrete
soliton. The DB is a strongly localized periodic regime of a lattice-based
system. The strong localization of the DB is often exponential, however,
in nonlinear systems it may even be hyper-exponential\cite{Flach2008}.
Simply put, the DB can be described as an oscillating envelope localized
around a single lattice site, or several sites in the case of a Multibreather
(MB). The DBs have been theoretically studied, and even experimentally
observed, in many fields of physics. Among the physical systems where
DBs have been encountered, are superconducting Josephson junctions\cite{Trias2000},
nonlinear magnetic metamaterials\cite{Lazarides2006}, electrical
lattices\cite{English2012}, micro-mechanical cantilever arrays\cite{Gutschmidt2010,Kimura2009a,Kenig2009,Sato2011,Sato2006},
Bose-Einstein condensates \cite{Trombettoni2001}, graphene sheets\cite{Fraile2016},
and chains of mechanical oscillators\cite{Gendelman2008,Cuevas2009,Gendelman2013,Grinberg2016,Perchikov2015}.

Either nonlinearity or disorder are essential features in order to
encounter the DBs, hence, exact analytic solutions are scarce. Theoretical
work on this phenomenon is primarily restricted to approximate techniques
and numerical study\cite{Flach1998,Flach2008,Cai1994,Romeo2016}.
Some known exceptions are the completely integrable Ablowitz-Ladik
model\cite{Ablowitz1976}, chains with homogeneous interaction \cite{Ovchinnikov1999}
and vibro-impact chains\cite{Gendelman2008,Perchikov2015,Gendelman2013,Grinberg2016}.
The latter has recently been expanded to asymmetric DBs in symmetric
and asymmetric settings\cite{grinberg2017}. Two former types of models --  
Ablowitz-Ladik and homogeneous interactions -- do not allow extension to higher dimensions. 
This Letter shows that the vibro-impact models can efficiently describe the DBs in a two-dimensional (2D) lattice.

We consider the DBs in the 2D lattice that comprises several identical
finite vibro-impact chains. Each chain's masses are linearly coupled
to their counterparts in the adjacent chains. The particles are allowed to move only in the directions of
the chains they belong to, so this system can be classified as the 2D scalar model.
Due to the complexity of the system, the treated model
is restricted to a conservative setup. The approach is based on the
refs. \cite{Gendelman2008,Perchikov2015,Gendelman2013,Grinberg2016,grinberg2017}, however the chain coupling does not allow exact analytical
solution in this case. The approach is based on generalized Fourier
series representation of the solution. An approximate solution is
obtained by truncation of the Fourier series and then solving the
resulting set of linear algebraic equations. Numerical study shows
a very rapid convergence of the series and therefore a very small
error. Furthermore, the nature of the model allows writing the monodromy
matrix explicitly, and therefore the linear stability of the DB and
MB solutions can be investigated without further approximation\cite{Strogatz1994}.
Stability analysis of the solutions reveals that pitchfork bifurcation
is possible despite the conservative symmetric model due to the coupling
of the chains.

As mentioned above, each particle is coupled by linear springs to its neighbors
in the chain, as well as to its counterparts in the adjacent chains.
Besides, each mass is subject to an identical on-site coupling \textendash{}
a linear spring with a symmetric pair of impact barriers located at
distances $u_{n,m}=\pm1$ from the trivial equilibrium position. This
unit scaling does not restrict the generality. The Hamiltonian of
the systems that includes $\left(M+1\right)$ chains of $\left(N+1\right)$
masses is written as follows:
\begin{equation}
\begin{array}{c}
H=\underset{m=0}{\overset{M}{\sum}}{\underset{n=0}{\overset{N}{\sum}}{\left(\cfrac{1}{2}p_{n,m}^{2}+V{\left(u_{n,m}\right)}\right)}}+\\
+\underset{m=0}{\overset{M}{\sum}}{\underset{n=0}{\overset{N-1}{\sum}}{W_{1}{\left(u_{n,m}-u_{n+1,m}\right)}}}+\\
+\underset{n=0}{\overset{N}{\sum}}{\underset{m=0}{\overset{M-1}{\sum}}{W_{2}{\left(u_{n,m}-u_{n,m+1}\right)}}}+\\
+\underset{m=0}{\overset{M}{\sum}}{W_{1}{\left(u_{N,m}-u_{0,m}\right)}}+\underset{n=0}{\overset{N}{\sum}}{W_{2}{\left(u_{n,M}-u_{n,0}\right)}},
\end{array}
\end{equation}
\begin{eqnarray}
V{\left(x\right)} & = & \begin{cases}
\cfrac{\gamma_{1}}{2}x^{2} & \,\,\,\,\,\left|x\right|<1\\
\mbox{infinity} & \,\,\,\,\,\left|x\right|=1
\end{cases},\\
W_{1}{\left(x\right)} & = & \cfrac{\gamma_{2}}{2}x^{2},\\
W_{2}{\left(x\right)} & = & \cfrac{\gamma_{3}}{2}x^{2},
\end{eqnarray}
where $p_{n,m}=\dot{u}_{n,m}$ is the momentum of each particle, $\gamma_{1}$,
$\gamma_{2}$ and $\gamma_{3}$ are the on-site coupling and chain
coupling stiffness coefficients respectively and $V{\left(x\right)}$,
$W_{1}{\left(x\right)}$ and $W_{2}{\left(x\right)}$ are the on-site,
coupling and chain coupling potentials, respectively. In principle,
the rigid on-site barriers without the linear anchoring spring are
sufficient to provide the existence of the DB solutions \cite{Gendelman2013};
in this work we consider more general case of $\gamma_{1}\geq0$.

We adopt here the traditional Newtonian model of the elastic impacts.
Namely, when at a certain time instance $t=t_{b}$ some particle achieves
the impact barrier ($u_{n}{\left(t_{b}\right)}=\pm1$), its velocity
is instantaneously modified according to the following law:

\begin{equation}
\dot{u}_{n}{\left(t_{b}+\right)}=-\dot{u}_{n}{\left(t_{b}-\right)}.\label{eq:7}
\end{equation}

We examine the conservative model where there is no external force
applied to the masses and no energy loss in the process, i.e. all impacts are elastic.
 The periodicity of the MB allows us to predict the time instances
of the collisions, and to express them as periodic external excitation
in the following manner:
\begin{equation}
\begin{array}{c}
\ddot{u}_{n,m}+\gamma_{1}u_{n,m}+\gamma_{2}\left(2u_{n,m}-u_{n+1,m}-u_{n-1,m}\right)+\\
+\gamma_{3}\left(2u_{n,m}-u_{n,m+1}-u_{n,m}\right)=2p_{n,m}\delta_{nk}\delta_{mr}\alpha{\left(t\right)},
\end{array}
\end{equation}
where $\alpha{\left(t\right)}=\sum_{j=-\infty}^{\infty}{\left(\delta{\left(t-\frac{\pi\left(2j+1\right)}{\omega}\right)}-\delta{\left(t-\frac{2\pi j}{\omega}\right)}\right)}$
describes the periodic impacts, $\delta{\left(t\right)}$ is the dirak
delta function, $\delta_{nm}$ is the Kronecker delta and, along with
$k$ and $r$, determines the localization sites, and $2p_{n,m}$
is the magnitude of the change of momentum during the impact. Since
the impacts are elastic $p_{n,m}$ determines the magnitude of the
velocity of the colliding mass just before, and right after the collision.
Additionally, finite chains with periodic boundary conditions are
considered, i.e. $u_{N+1,m}{\left(t\right)}=u_{0,m}{\left(t\right)}$
and $u_{n,M+1}{\left(t\right)}=u_{n,0}{\left(t\right)}$ where $N+1$
is the number masses in each chain and $M+1$ is the number of chains.
Without loss of generality, we let the mass denoted by $m=0$ and
$n=0$ to be a localization site. Note, however, that at least one site cannot be a localization site in order to excludes a trivial case, when all sites in the lattice are excited.

The terms for the periodic impacts can also be written in the form of generalized
Fourier series:
\begin{equation}
\alpha{\left(t\right)}=-\frac{2\omega}{\pi}\sum_{j=0}^{\infty}{\cos{\left(\left(2j+1\right)\omega t\right)}}.
\end{equation}

The solution can now also be written as a generalized Fourier series in a similar form:
\begin{equation}
u_{n,m}=\sum_{j=0}^{\infty}{u_{n,m,j}\cos{\left(\left(2j+1\right)\omega t\right)}}.
\end{equation}

Plugging into the equations of motion and separating to different
harmonics yields :
\begin{equation}
\begin{array}{c}
-\left(2j+1\right)^{2}\omega^{2}u_{n,m,j}+\gamma_{1}u_{n,m,j}+\\
+\gamma_{2}\left(2u_{n,m,j}-u_{n+1,m,j}-u_{n-1,m,j}\right)+\\
+\gamma_{3}\left(2u_{n,m,j}-u_{n,m+1,j}-u_{n,m,j}\right)=-\frac{4\omega p_{n,m}}{\pi}\delta_{nk}\delta_{mr}.
\end{array}\label{eq:13}
\end{equation}

Additionally to this infinite set of equation, there is also the impact
condition for each impacting mass:
\begin{equation}
u_{k_{s},r_{s}}{\left(0\right)}=\sum_{j=0}^{\infty}{\left(-1\right)^{k_{s}+r_{s}}u_{k_{s},r_{s},j}}=\pm1,\label{eq:17}
\end{equation}
where the $\pm$ sign determines whether the oscillations of the specific
site is in or out-of phase with respect to the other localization
sites.

In order to obtain an approximate solution, we limit the number of
harmonics considered to the first $L+1$ . Thus, we obtain a set of
$\left(L+1\right)\left(N+1\right)\left(M+1\right)+\left(\mbox{Number of impacting masses}\right)$
linear equations which can be solved in order to find an approximate
solution. It is important to note that if a solutions exists, convergence
is certain and $L$ determines the accuracy of the obtained approximation.

We investigate the stability of the MB solutions with the help of
Floquet theory\cite{Strogatz1994}. Since the explored model allows
explicit construction of the monodromy matrix, it is easy to find
its eigenvalues for every set of parameters. Thus, broad regions of
the parameter space can be explored for various structures of the
breathers with minimal numerical efforts. Moreover, the eigenvectors
corresponding to the unstable Floquet multipliers can be easily computed
and examined to give a qualitative insight into physical mechanisms
of the loss of stability.

The governing equations of motion for can be re-written in the following
equivalent form:
\begin{equation}
\dot{\vec{u}}=\mbox{A}\vec{u},
\end{equation}

where,
\begin{equation}
\vec{u}=\left[\begin{array}{c}
\vec{x}\\
\dot{\vec{x}}
\end{array}\right]
\end{equation}
\begin{equation}
\vec{x}=\left[\begin{array}{cccc}
u_{00} & \cdots & u_{\left(N-1\right)M} & u_{NM}\end{array}\right]^{T}
\end{equation}
 and:
\begin{equation}
\mbox{A}=\left[\begin{array}{cc}
\mbox{0} & \mbox{I}\\
\tilde{\mbox{A}} & \mbox{0}
\end{array}\right]_{2\left(N+1\right)\left(M+1\right)\times2\left(N+1\right)\left(M+1\right)},
\end{equation}
\begin{equation}
\tilde{\mbox{A}}=\left[\begin{array}{cccccc}
\mbox{A}_{1} & \mbox{A}_{2} & 0 & \cdots & 0 & \mbox{A}_{2}\\
\mbox{A}_{2} & \mbox{A}_{1} & \mbox{A}_{2} & 0 & \cdots & 0\\
0 & \mbox{A}_{2} & \ddots & \ddots & \ddots & \vdots\\
\vdots & \ddots & \ddots & \mbox{A}_{1} & \mbox{A}_{2} & 0\\
0 & \cdots & 0 & \mbox{A}_{2} & \mbox{A}_{1} & \mbox{A}_{2}\\
\mbox{A}_{2} & 0 & \cdots & 0 & \mbox{A}_{2} & \mbox{A}_{1}
\end{array}\right]_{\begin{array}{c}
\left(N+1\right)\left(M+1\right)\times\\
\times\left(N+1\right)\left(M+1\right)
\end{array}},
\end{equation}
\begin{equation}
\mbox{A}_{1}=\left[\begin{array}{cccccc}
\sigma & -\gamma_{2} & 0 & \cdots & 0 & -\gamma_{2}\\
-\gamma_{2} & \sigma & -\gamma_{2} & 0 & \cdots & 0\\
0 & -\gamma_{2} & \ddots & \ddots & \ddots & \vdots\\
\vdots & \ddots & \ddots & \sigma & -\gamma_{2} & 0\\
0 & \cdots & 0 & -\gamma_{2} & \sigma & -\gamma_{2}\\
-\gamma_{2} & 0 & \cdots & 0 & -\gamma_{2} & \sigma
\end{array}\right]_{\left(N+1\right)\times\left(N+1\right)},
\end{equation}

\begin{equation}
\mbox{A}_{2}=-\gamma_{3}\mbox{I}_{\left(N+1\right)\times\left(N+1\right)},
\end{equation}
where $\sigma=\gamma_{1}+2\gamma_{2}+2\gamma_{3}$

Here $\tilde{A}$ is the Laplace adjacency matrix of the system. For the
forced-damped model, minor modification is required:
\begin{equation}
\dot{\vec{v}}=\mbox{A}\vec{v}+\vec{F},
\end{equation}
where $\vec{F}=F{\left(t\right)}\left[\begin{array}{ccc}
0 & \cdots & 0\end{array}\begin{array}{ccc}
1 & \cdots & 1\end{array}\right]^{T}$.

All considered solutions
are symmetric in a sense that the successive impacts for each particle
are divided by half-period intervals, and absolute amounts of momentum
transferred to given particle in the course of given impact is the
same. From the above equation, it is easy to derive the matrix, that
describes the evolution of the perturbed phase trajectory between
two successive impacts:
\begin{equation}
\mbox{L}=\exp\left(\cfrac{\pi}{\omega}\mbox{A}\right).
\end{equation}

To describe the evolution of the perturbed phase trajectory in the
course of impacts, we apply a formalism of the saltation matrix\cite{Fredriksson2000}.
Since the impacts are instantaneous independent events, they can be
treated separately and then combined to result in the following saltation
matrix:
\begin{equation}
\mbox{S}=\left[\begin{array}{cc}
\mbox{\ensuremath{\tilde{\mbox{S}}}} & \mbox{0}\\
\hat{\mbox{S}} & \mbox{\ensuremath{\tilde{\mbox{S}}}}
\end{array}\right]_{2\left(N+1\right)\left(M+1\right)\times2\left(N+1\right)\left(M+1\right)},
\end{equation}

where,{\small{}
\begin{equation}
\tilde{\mbox{S}}=\left[\begin{array}{ccccc}
\epsilon_{00} & 0 & \cdots & \cdots & 0\\
0 & \epsilon_{10} & 0 &  & \vdots\\
\vdots & 0 & \ddots & \ddots & \vdots\\
\vdots & \mbox{} & \ddots & \epsilon_{\left(N-1\right)M} & 0\\
0 & \cdots & \cdots & 0 & 1
\end{array}\right]_{\begin{array}{c}
\left(N+1\right)\left(M+1\right)\times\\
\times\left(N+1\right)\left(M+1\right)
\end{array}},\label{eq:53}
\end{equation}
\begin{equation}
\epsilon_{ij}=1-\left(1+e\right){\delta_{ik}\delta_{jr}}
\end{equation}
\begin{equation}
\hat{\mbox{S}}=\left[\begin{array}{ccccc}
s_{00} & 0 & \cdots & \cdots & 0\\
0 & s_{10} & 0 &  & \vdots\\
\vdots & 0 & \ddots & \ddots & \vdots\\
\vdots &  & \ddots & s_{\left(N-1\right)M} & 0\\
0 & \cdots & \cdots & 0 & 0
\end{array}\right]_{\begin{array}{c}
\left(N+1\right)\left(M+1\right)\times\\
\times\left(N+1\right)\left(M+1\right)
\end{array}}.\label{eq:54}
\end{equation}
\begin{equation}
s_{ij}=\cfrac{\left(1+e\right){\delta_{ik}\delta_{jr}\ddot{u}_{i,j}{\left(\phi-\right)}}}{p_{i,j}}
\end{equation}
}{\small \par}

Due to the symmetry of the even functions composing the Fourier series,
the monodromy matrix can be written compactly as follows:
\begin{equation}
\mbox{M}=\left(\mbox{L}\mbox{S}\right)^{2}.
\end{equation}

As mentioned above, the eigenvalues of this monodromy matrix are computed
numerically for the given parameter values. The resulting stability
pattern in the space of parameters are exemplified in the next section.

\begin{figure}[h]
\begin{centering}
\includegraphics[width=1\columnwidth]{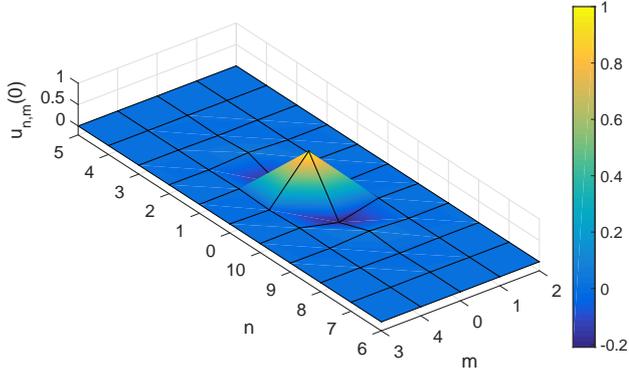}
\par\end{centering}
\caption{\label{fig1}The frame of a DB for $\gamma_{2}=0.2$, i.e. the displacements of the masses at the instance of the impact.}
\end{figure}

\begin{figure}[h]
\begin{centering}
\includegraphics[width=1\columnwidth]{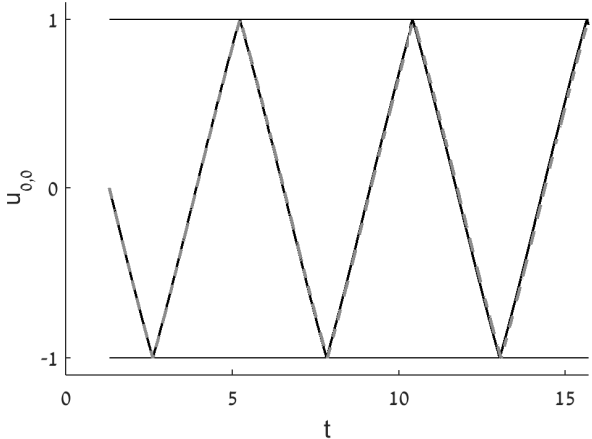}
\par\end{centering}
\caption{\label{fig2}Displacement of one of the impacting masses for a multibreather
with two localization sites at $\left(0,0\right)$ and $\left(3,2\right)$,
i.e. $n=3$ and $m=2$. Dashed gray line corresponds to the analytical
result and solid black line corresponds to numerical result.}
\end{figure}

An example of the DB shape i.e. the displacements of the masses at the
time instance of the impact, is presented in fig. \ref{fig1}, where one can
appreciate already the strong localization of the solution. For qualitative
understanding of the properties of the analytically obtained DB solution,
and in order to verify the accuracy of the numerical algorithms, we compare the
results of the analysis to numerical simulations. The simulations were
performed in MatLab; the vibro-impact was modeled according to the
impact law using built-in event-driven algorithms with Runge-Kutta
(RK) solver. The results of the numerical simulations were in good agreement
with the analytic solution despite the approximation as illustrated
in fig. \ref{fig2}. It is important to note that the number of masses
in these simulations is relatively large; hence, the number of leading
harmonics chosen here is not very large to maintain a reasonable computation
time. Numerical examination shows great improvement in accuracy when
further increasing the number of harmonics at the cost of a much longer
simulation.

Unless stated otherwise, the parameters chosen for all simulations
are:
\begin{equation}
\begin{array}{c}
\begin{array}{ccc}
N=10, & M=4, & \omega=1.2,\end{array}\\
\begin{array}{ccc}
\gamma_{1}=0.1, & \gamma_{2}=0.01, & \gamma_{3}=0.02.\end{array}
\end{array}
\end{equation}

\begin{figure}[h]
\begin{centering}
\includegraphics[width=1\columnwidth,trim= 8 1 15 10, clip=true]{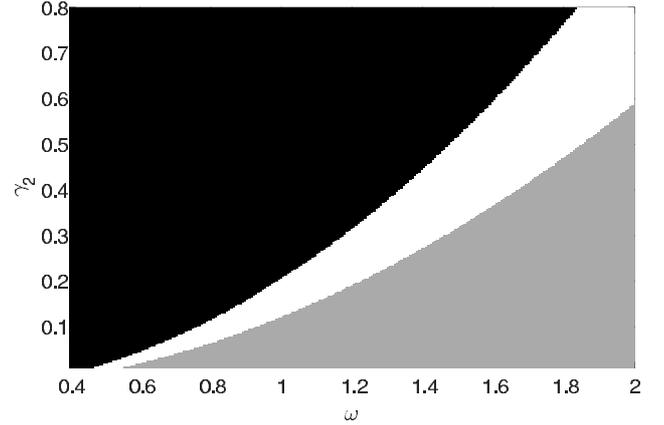}
\par\end{centering}
\caption{\label{fig3}Stability map in the frequency-coupling stiffness plane
for a multibreather with two localization sites at $\left(0,0\right)$
and $\left(3,2\right)$, i.e. $n=3$ and $m=2$ is the second localization
site. Gray region denotes stable solutions and white region denotes
unstable solutions. In the black region the solution with the considered
structure does not exist.}
\end{figure}

\begin{figure}[h]
\begin{centering}
\includegraphics[width=1\columnwidth]{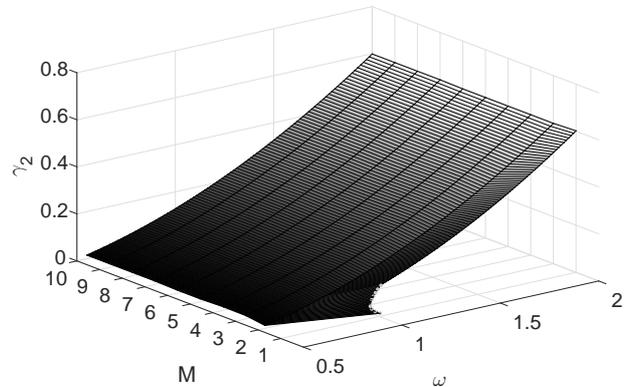}
\par\end{centering}
\caption{\label{fig4}Stability threshold as a function of the number of chains
for a DB. The white dot marking corresponds to stability loss via
pitchfork bifurcation in oppose to The Neimark-Sacker bifurcation
elsewhere.}
\end{figure}

\begin{figure}[h]
\begin{centering}
\includegraphics[width=1\columnwidth]{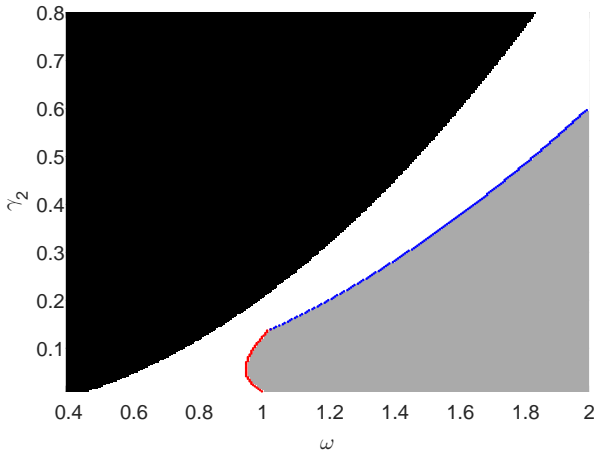}
\par\end{centering}
\caption{\label{fig5}Stability map in the frequency-coupling stiffness plane
for a DB. Gray region denotes stable solutions and white region denotes
unstable solutions. In the black region the solution with the considered
structure does not exist. The Red and blue marking corresponds to
stability loss via pitchfork and Neimark-Sacker bifurcation, respectively.}
\end{figure}

Figure \ref{fig3} shows a typical existence-stability map for the
MB. Similarly to previous work, the mechanism for loss of stability
discovered in this example of a conservative MB is the Neimark Sacker
Bifurcation. In fig. \ref{fig4} we demonstrate weak correlation between
the stability threshold and the number of chains except for a very
small number of chains. This weak correlation is explained by the
strong localization of the solution. Note that this may also be interpreted
as the correlation between the stability threshold and the number
of particles in the chain as the equations of motion shows these are
interchangeable. Another interesting result is in the case of $M=1$,
i.e. two coupled chains, where we can see the threshold differs in
the lower frequencies. At lower frequencies the dashed marking in
fig. \ref{fig4} corresponds to pitchfork bifurcation, that is previously
not encountered for a conservative DB. For clarification, the difference
for $M=1$ is seen in fig. \ref{fig5}.

\begin{figure}[h]
\begin{centering}
\includegraphics[width=1\columnwidth]{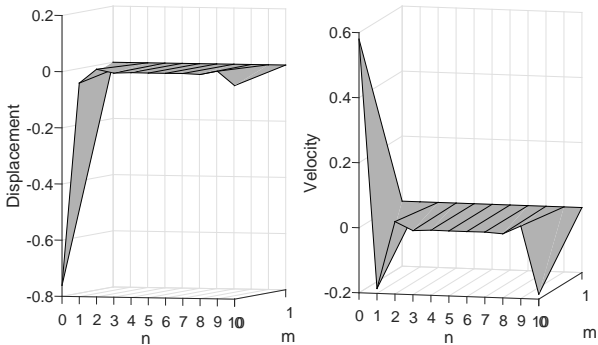}
\par\end{centering}
\caption{\label{fig6}Terms corresponding to displacement (left) and velocity
(right) of an eigenvector associated with the pitchfork bifurcation. }
\end{figure}

Figure \ref{fig6} shows the eigenvectors corresponding to the pitchfork bifurcation.
While it is difficult to show numerically, we can learn from the strongly
localized form of the eigenvectors that it is likely to lead to breaking
of symmetry as often associated with the pitchfork bifurcation. This
is an interesting and somewhat surprising finding as the calculations
in ref. \cite{grinberg2017} clearly shows that an asymmetric DB is not possible
for the single conservative chain.

To conclude, in this work, we derive the approximate DB and MB solutions
for coupled vibro-impact chains.  Contrary to the single vibro-impact chain,
an exact solution cannot be simply derived. However, similarly to
the single chain, the monodromy matrix can be written explicitly,
and one can analyze the stability of the approximate solutions without
further loss of accuracy.

Another interesting finding is the appearance of the pitchfork bifurcation.
While it was proven analytically in ref. \cite{grinberg2017} that symmetry breaking
is not possible for the single chain in a symmetric conservative model,
stability analysis of the DB solution shows that it is possible in
coupled chains.

Further investigation is needed in two aspects. The first is numerical
validation of the symmetry breaking related  to the pitchfork bifurcation.
This may prove to be of some complexity as the solutions are not attractors.
The second challenge is a derivation of the approximate solution, when external
forcing and damping are included. The difficulty lies with the addition of nonlinearity
to the algebraic equations that has to be solved and may cost in a
less accurate approximation and larger computational effort.

The authors are grateful to Israel Science Foundation (grant 838/13) for financial support.

\bibliographystyle{plain}
\bibliography{bibtex2}

\end{document}